\documentstyle[12pt]{article}

\begin{document}
\newcommand {\be}{\begin{equation}}
\newcommand {\ee}{\end{equation}}
\newcommand {\bea}{\begin{array}}
\newcommand {\cl}{\centerline}
\newcommand {\eea}{\end{array}}
\def\pr{\partial}
\def \a'{\alpha'}
\baselineskip 0.65 cm
\begin{flushright}
IPM/P-98/22 \\
hep-th/9810179
\end{flushright}

\begin{center}
{\Large{\bf Super Yang-Mills Theory on Noncommutative Torus from Open Strings 
Interactions}}
\vskip .5cm

M.M. Sheikh-Jabbari
\footnote{ E-mail: jabbari@theory.ipm.ac.ir } \\

\vskip .5cm

 {\it Institute for studies in theoretical Physics and Mathematics 
IPM,

 P.O.Box 19395-5531, Tehran, Iran}\\
\end{center}

\vskip 2cm
\begin{abstract}
Considering the scattering of massless open strings attached to a D2-brane
living in the $B$ field background, we show that corresponding scattering 
upto the order of $\a'^2$ is exactly given by the gauge theory on noncommutative
background, which is characterized by the Moyal bracket.

\end{abstract}
\newpage
{\it Introduction}

After the novel work of Witten [1], it was realized that the low energy 
dynamics of open strings attached to a D-brane is governed by the Super Yang-Mills
(SYM) theory defined on the D-brane world volume. The gauge group of the 
corresponding SYM theory is $U(N)$, $N$ being the number of coincident D-branes.
Elaborating on the point that such open strings describe the dynamics of 
D-branes [2], led to the M(atrix)-model [3].
M(atrix)-model as a conjecture for discrete light cone quantization (DLCQ) 
of M-theory should also contain the 
background three-form field of 11 dimensional supergravity, such a 
background field was studied in [4]. There Connes, Douglas and Schwarz (CDS)
conjectured that the SYM theory on a noncommutative torus properly describes
the three-form field background.

Since the eleven dimensional three-form is related to the NSNS two from field
of string theory, the CDS conjecture means that the dynamics of D0-branes, or
more generally any other D-brane, in a $B$ filed background is described by
a gauge theory living on a noncommutative torus, $T^d_{\theta}$. It has been 
argued that the $\theta$ parameter of the torus should be identified with
the constant $B$ field background [5]. Thereafter the D-brane dynamics or the
M(atrix)-model on a noncommutative space has been vastly studied [6,7,8,9,10,  
11,12,13,14,15,16,17]. It was argued in [5] that, we can realize the characteristics 
of the SYM on noncommutative torus, the Moyal bracket, in string theory.
In [9,13] extending the ideas of [5], they studied the D0-brane dynamics in 
both constant and non-constant B 
field background and showed that this will modify the ordinary multiplication
of field with the Moyal product. 

In this paper using the usual string theory methods, we study the interaction
of open strings attached to a D2-brane living in the B field background. 
We show that the same scattering amplitudes upto second order of $\alpha'$,
can be obtained from the SYM theory on noncommutative torus. We  show that
$O(\alpha')$ corrections to SYM are cancelled at super open string tree level.
By going to a dual picture of above, i.e. a D-string in B field background
considering the open string scattering, we will give an intuitive way of 
finding the Fourier transformed of the interaction terms of the effective
action. 

{\it Open string scattering}
\newline
In order to calculate the open strings scattering amplitude, first we discuss
their mode expansions. Although the following argument holds for the most
general case, for simplicity let us consider a D2-brane winding around an 
orthogonal torus defined by:
\be
\tau={iR_2 \over R_1}\;\;\;\; , \;\;\;\;\ \rho=iR_1R_2+B.
\ee
The open string attached to such a D2-brane satisfy the boundary conditions 
[18]

\be
\left\{  \begin{array}{cc}
\pr_{\sigma}X^0=0 \\
\partial_{\sigma}X^{i}+B^i_j \partial_{\tau}X^j=0  \;\;\; i,j=1,2\\
\pr_{\tau} X^a=0 \;\;\;\ ,\;\;\; a=3,...,9.
\end{array}\right.
\ee
Hence the mode expansion of these open strings are [15]

\be
\left\{  \begin{array}{cc}
X^0=x^0+ p^0\tau+ \sum_{n\neq 0}a^0_n{e^{-in\tau} \over n}\cos n\sigma\\
X^{i}=x^i+(p^i \tau-B^i_j p^j \sigma)+ \sum_{n\neq 0} {e^{-in\tau} \over n}
\bigl(ia^i_n \cos n\sigma + B^i_j a^j_n \sin n\sigma \bigr) \;\;\;\ i,j=1,2\\
X^a=x^a+ \sum_{n\neq 0}a^a_n {e^{-in\tau} \over n}\sin n\sigma ,
\end{array}\right.
\ee
where $x^i$ show an arbitrary point on D2-brane and $p^i$ are vectors
on the dual torus:
\be
p^i={n_i \over R_i} \;\;\; , n_i \in Z.
\ee
We observe that in the case of non-zero B field the coefficient of $\sigma$
in the mode expansion is proportional to $p^j$. As it is shown in [15], this
is the root of having noncommutative coordinates:
\be
[x^{i},p^{j}]=i\eta^{\mu\nu} \;\;\ , \;\;\; [x^i,x^j]=-2i B^{ij}.
\ee
This is a clue showing the field theory living on a D2-brane in B field  
background, which is described by these open strings, is not a conventional
local field theory.

To find the effective low energy theory governing these open strings, we 
consider their scattering processes.

{\it Three super open string scattering}
\newline
Let us consider three massless vector vertex [19]:
\be
S_{int3}=<V_1V_2V_3>,
\ee
where 
\be 
V_a=\int d\sigma \xi_a.\Pi exp(ik_a.X(\sigma,\tau)),
\ee
with $\xi_a, k_a$ are the polarization and momentum of the vector state:
$\xi_a.k_a=k_a^2=0$. Here we only consider the nontrivial cases, i.e.
$\xi_a, k_a$ are vectors on the torus. 
$\Pi$ is the conjugate momentum of $X(\sigma,\tau)$,
which is proportional to $p^i$ [20], minus a fermionic part which is
${-1 \over 4}\gamma_{ab}^{ij}S^aS^bk^j$ [19] and $X$ is the mode
expansion given by eq.(3).
\newline
Calculating $S_{int3}$ with string theory methods [19], we find
\be
S_{int3}=g \xi_1^{\mu}\xi_2^{\nu}\xi_3^{\rho}\ t_{\mu\nu\rho} 
exp(\sum_{a<b} iB_{ij}k^i_ak^j_b)\delta(\sum_{a} k_a^{i}),
\ee
where 
\be
t_{\mu\nu\rho}=k_{2\rho}\eta_{\mu\nu}+k_{3\nu}\eta_{\mu\rho}+
k_{1\mu}\eta_{\rho\nu}.
\ee
The term $exp(\sum_{a<b} iB_{ij}k^i_ak^j_b)$, which shows the B dependence of
the interaction, considering the momentum conservation, can be written as
$$
exp(iB_{ij}k^i_ak^j_b) \;\;\;\ a<b.
$$
So $S_{int3}$ corresponds to the cubic interaction derived from the $tr F^2$
in SYM on noncommutative torus. More explicitly, $S_3$ shows the Fourier 
transformed from of 
\be
\bea{cc}
\int d^3x \pr_{\mu}A_{\nu}(x) exp(iB{\pr \over \pr x'^1}{\pr \over \pr x'^2})
A_{\mu}(x'^1,x^2)A_{\nu}(x^1,x'^2)|_{x'^i=x^i}\\
=\int d^3x \pr_{\mu}A_{\nu}(x) \ [A_{\mu},A_{\nu}]_{M.B.}.
\eea\ee
{\it Four open string scattering}
\newline
To obtain the full effective action, we should also consider the $M$ open string
scattering, but it is easy to show that all the higher M $(M>4)$ are
reduced to the three and four open string scattering. Hence we consider the four 
string massless vector vertex:
\be
S_{int4}=<V_1V_2V_3V_4>.
\ee
Again by the usual string theory calculations: 
\be
S_{int4}={-1 \over 2}g^2 \xi_1^{\mu}\xi_2^{\nu}\xi_3^{\alpha}
\xi_4^{\beta}\ K_{\mu\nu\alpha\beta} 
exp(\sum_{a<b} iB_{ij}k^i_ak^j_b)\delta(\sum_{a} k_a^{i}),
\ee
where the kinematic factor $K$, is symmetric under the exchange of
1,2 with 3,4, and antisymmetric under exchange of 1 with 2 and 3 with 4, 
and contains two type of terms: The term proportional to ${-1\over 4}
\eta_{\mu\nu}\eta_{\alpha\beta}$ and its permutations, and the terms of the 
form ${1\over 2} k_a^{\mu}k_b^{\alpha}\eta_{\nu\beta}$. 
Since \footnote{ All other permutations in result is also possible.}
$$
exp(\sum_{a<b} iB_{ij}k^i_ak^j_b)=exp(iB_{ij}k^i_1k^j_2)exp(iB_{ij}k^i_3k^j_4)
$$
Taking only the massless poles in $K$, the first type gives rise to the
terms 
\be 
[A_{\mu},A_{\nu}]_{M.B.}[A^{\mu},A^{\nu}]_{M.B.},
\ee
in the effective action, and the second type, is described  by
\be
\pr_{\mu}A_{\nu}\ [A_{\mu},A_{\alpha}]_{M.B.}\pr_{\alpha}A_{\nu}.
\ee
As we see in the tree level of the super open string scattering there is 
no $O(\alpha')$ corrections to the effective action of SYM on noncommutative 
torus.

Putting the results of three and four super open string scattering together,
the low energy effective action reads as:
\be
S=\int trF^2 d^3x,
\ee
with
\be
F_{\mu\nu}=\pr_{\mu}A_{\nu}-\pr_{\nu}A_{\mu}+[A_{\mu},A_{\nu}]_{M.B.}.
\ee
and the $tr$ is performed on the noncommutative torus.

{\it Intuitive calculations}
\newline
The calculations given above can also be understood in an intuitive way. To
do so, we consider a D-string wound around the cycle $R_1$, with a B field 
background. We will momentarily show that generalizing the ideas of [5], one
can obtain the full SYM on $T^2_{\theta}$.
As we will see in the D-string version, we should perform a Fourier transformation  
to obtain the Moyal bracket structure, i.e. the T-duality in string theory is
like the Fourier expansion in related field theory.

In three open string scattering, two open strings can interact when the end 
of the first coincides with the beginning of the second. 
It is worth noting that the open strings attached to D-string are oriented.
Here we deal with three open strings attached to such a D-string.
Mode expansion of these open strings are [15] 

\be\left\{ \bea{cc}
X^i=x_0^i + p^i \tau+ L^i \sigma + Oscil. \;\;\; , i=1,2  \\
X^0=x^0_0+p^0\tau+ Oscil. \\
X^a=x^a_0+ Oscil. \;\;\; , \;\;\; a=3,...,9 
\eea\right.
\ee
where 

\be\bea{cc} 
p^1=\ {r \over R_1}  \;\;\;\;\;\;\ , \;\;\;\ p^2=0 \\
L^1=\ {qB \over R_1} \;\;\;\;\;\;\ , \;\;\;\ L^2=qR_2 \;\;\ r,q\in Z.
\eea\ee
Hence the beginning and the end of these open strings upto the torus 
identifications are
\be
\bea{cc}
{\rm beginning} \\ {\rm end}
\eea
(1)\left\{\bea{cc} 
(x^1,0) \\ (x^1+l^1,0) \;\;\;\;
\eea\right.
(2)\left\{\bea{cc}
(x^2,0) \\ (x^2+l^2,0) \;\;\;\;
\eea\right.
(3)\left\{\bea{cc} 
(x^3,0) \\ (x^3-l^3,0)
\eea\right.
\ee
with  $l^i={q^iB \over R_1}$.
The interaction condition reads as:
\be
x^1+l^1=x^2 \;\;\;\;\ ,\ \;\;\;\ x^2+l^2=x^3-l^3 \;\;\; ,\;\;\ x^3=x^1.
\ee
Resulting in:
\be
l^1+l^2+l^3=0
\ee
Whenever the field associated with the open string, aside from its tensorial
structure, is denoted by $\Phi(x,l)$, the interaction of three open string is 
given by:
\be\bea{cc}
S_{int3}=\sum_{q^i} \int dx^1dx^2dx^3 \Phi_1(x^1,l^1) \Phi_2(x^2,l^2)
\Phi_3(x^3,l^3)\ \delta(x^3-x^1) \delta(x^2-x^1-l^1)\delta(q^1+q^2+q^3) \\ 
=\sum_{q^1,q^2}\int dx^1 \Phi_1(x^1,q^1) \Phi_2(x^1+l^1,q^2) 
\Phi_3(x^1,-q^1-q^2).
\eea\ee
After Fourier transformation and putting $\sigma_i=R_ix^i$ [5]
\be
S_{int3}=\int d\sigma^1\ d\sigma^2 \Phi_1(\sigma^1,\sigma^2)
[\Phi_2,\Phi_3]_{_{M.B.}}(\sigma^1,\sigma^2).
\ee
We can also have the four open string scattering. Let us consider two
open strings as:
\be
\bea{cc}
{\rm beginning} \\ {\rm end}
\eea
(1)\left\{\bea{cc} 
(x^1,0) \\ (x^1+l^1,0) \;\;\;\;
\eea\right.
(2)\left\{\bea{cc}
(x^2,0) \\ (x^2+l^2,0) \;\;\;\;
\eea\right.
\ee
These open strings can split at some mid-point and join again, so 
that after scattering we have two open strings starting at $(x^1,0)$ and
$(x^2,0)$ and ending at $(x^2+l^2,0)$ and $(x^1+l^1,0)$ respectively:
\be
\bea{cc}
{\rm beginning} \\ {\rm end}
\eea
(4)\left\{\bea{cc} 
(x^1,0) \\ (x^2+l^2,0) \;\;\;\;
\eea\right.
(3)\left\{\bea{cc}
(x^2,0) \\ (x^1+l^1,0) \;\;\;\;
\eea\right.
\ee
\newline
Imposing the condition that these open strings should end on the D-string forces
\be
x^1+l^1-x^2=l,
\ee
where $l$ is of the form of ${nB \over R_1}$ for arbitrary integer $n$. Hence 
the effective interaction can be written as
\be
S_{int4}=\sum_{q^1,q^2,n} \int dx^1dx^2 \Phi_1(x^1,l^1) \Phi_2(x^2,l^2)
\Phi_3(x^1,l^1-l)\ \Phi_4(x^2,l). 
\ee
Then one can Fourier transform eq.(27) to obtain:
\be\bea{cc}
S_{int4}=\int d\sigma^1\ d\sigma^2 exp(iB{\pr \over \pr\eta^1} 
{\pr \over \pr\eta^2})\Phi_1(\sigma^1,\eta^2)\Phi_2(\eta^1,\sigma^2)|_
{\eta^i=\sigma^i}\ exp(iB{\pr \over \pr\xi^1} 
{\pr \over \pr\xi^2})\Phi_3(\sigma^1,\xi^2)\Phi_4(\xi^1,\sigma^2)|_
{\xi^i=\sigma^i}\\
=\int d\sigma^1\ d\sigma^2 [\Phi_1,\Phi_2]_{M.B.}
[\Phi_3,\Phi_4]_{M.B.}.
\eea\ee
So the interaction part of the action is:
\be
S_{int}=S_{int3}+S_{int4}=\int d\sigma^1\ d\sigma^2 
\bigl(\Phi_1\ [\Phi_2,\Phi_3]_{M.B.}+
[\Phi_1,\Phi_2]_{M.B.}[\Phi_3,\Phi_4]_{M.B.}\bigr).
\ee
In our case we know that the $\Phi$ fields describing the light states of
open strings are vector states, so eq.(29) gives the interaction part of the 
SYM on $T^2_{\theta}$, with $\theta$ identified with $B$.

{\it Discussion}
\newline
In this paper considering three and four open strings scattering, we explicitly
showed that upto tree open string level at low energies, the SYM on 
$T^2_{\theta}$ effectively describe the massless vector open strings.

Appearance of the phase factor in eq.(8) and eq.(12) is a result of using
mixed boundary conditions, the eq.(2). This phase factor first pointed out  
in [22]. The background field considered there ([22]) was the electric, which
as discussed in [15,20] will not lead to any noncommutativity in space-time.

We should note that considering the {\it super} strings removes the 
$\alpha'$ corrections to the effective action at tree level and hence
the first correction to the SYM is of the order of $\alpha'^2$. 
These correction terms as calculated in [21], are 
\be
L={-1\over 4} trF^2 +{1\over 8}(2\pi\alpha')^2(F^4-{1\over4}(F^2)^2)+
O(\alpha'^3),
\ee
but in our case, i.e. when we have a non-zero background B field,  
$F_{\mu\nu}=\pr_{[\mu}A_{\nu]}(x) + [A_{\mu},A_{\nu}]_{M.B.}$.

This correction is in perfect agreement with the leading terms appearing
in the expansion of DBI action on $T^2_{\theta}$.
So it seems that one can regain the full DBI action by considering
all order corrections to the SYM theory.

In this paper, we studied scattering amplitudes of open strings attached to    
a D-brane with the B field turned on and verified that they are consistent
with the SYM theory on noncommutative torus. It would be interesting to
generalize the calculations of [23], the N coincident D-branes, to the case  
with non-vanishing B field background. In this case we expect to see both
the Moyal and the gauge group commutators in the effective action. 

Obtaining the SYM on noncommutative torus from the open strings scattering,  
tells us that this theory, although being a non-local field theory, may be 
renormalizable. More over we can show that this gauge theory can be 
obtained from a dual gauge theory on a commutative background [24].
In this way, one can hope to handle the question of renomalizability 
of SYM on noncommutative torus.

{\bf Acknowledgements}

I would like to thank H. Arfaei for fruitful comments. I also thank 
F. Ardalan,  K. Kaviani and M. Garousi for helpful discussions.


\begin{thebibliography}{99}
\bibitem{}  E. Witten, {\it Nucl. Phys.} {\bf B460} (1996) 335.

\bibitem{} J. Polchinski, {\it Phys. Rev. Lett.} 75(1995) 4727.


\bibitem{}  T. Banks, W. Fischler, S.H. Shenker, and L. Susskind, 
Phys. Rev.{\bf D55} (1997) 5112,

N. Ishibashi, H. Kawai, Y. Kitazawa and A. Tsuchiya, {\it Nucl. Phys.} 
{\bf B498} (1997) 467. 

L. Susskind, "Another Conjecture about M(atrix) Theory", hep-th/9704080.

\bibitem{}
A. Connes, M.R. Douglas, A. Schwarz,
"Noncommutative Geometry and Matrix Theory: Compactification on Tori",
hep-th/9711162.

\bibitem{} M. R. Douglas, C. Hull, "D-branes and Noncommutative Torus",
hep-th/9711165.

\bibitem{} P.-M. Ho, Y.-S. Wu,
"Noncommutative Gauge Theories in Matrix Theory", hep-th/9801147.

\bibitem{} M. Li,
"Comments on Supersymmetric Yang-Mills Theory on a Noncommutative Torus",
hep-th/9802052.

\bibitem{}
A. Schwarz, "Morita equivalence and Dualities",  hep-th/9805034.

\bibitem{}
Y.-K. E. Cheung, M. Krogh,
"Noncommutative Geometry From $0$-Branes in a Background $B$ Field",
hep-th/9803031.

\bibitem{}  
T. Kawano and K. Okuyama,  "Matrix Theory on Noncommutative Torus", 
hep-th/9803044. 

\bibitem{} F. Ardalan, H. Arfaei, M. M. Sheikh-Jabbari,
"Mixed Branes and M(atrix) Theory on Noncommutative Torus", hep-th/9803067.

\bibitem{}
P.-M. Ho, "Twisted Bundle on Quantum Torus and BPS States in Matrix Theory",
hep-th/9803166.

\bibitem{}
H. Garcia-Compean, "On the Deformation Quantization Description of 
Matrix Compactifications", hep-th/9804188.

\bibitem{}
B. Morariu, B. Zumino, "Super Yang-Mills on the Noncommutative Torus",
hep-th/9807198,

\bibitem{} F. Ardalan, H. Arfaei, M. M. Sheikh-Jabbari,
"Noncommutative Geometry form Strings and Branes", hep-th/9810072.

\bibitem{}
D. Brace, B. Morariu, B. Zumino, "Dualities of the Matrix Model From T-duality
of the Type II Strings", hep-th/9810099.

\bibitem{}
C. Hofman and E. Verlinde, "U-duality of Born-Infeld on the Noncommutative  
Two Torus", hep-th/9810116.

\bibitem{}  R. G. Leigh, {\it Mod. Phys. Lett.} {\bf A4} 28 (1989) 2767. 
 
\bibitem{}
M.B. Green, J.H. Schwarz, E. Witten, "Super String Theory", Vol. 1, chap. 7.

\bibitem{}  H. Arfaei and M.M. Sheikh-Jabbari, hep-th/9709054, 
{\it Nucl. Phys.} {\bf B526} (1998) 278.  

\bibitem{}
A.A. Tseytlin, {\it Nucl. Phys.} {\bf B276} (1986) 391.

\bibitem{}
S. Gukov, I. Klebanov, A. Polyakov, "Dynamics of (n,1) strings", hep-th/9711112,
{\it Phys. Lett.} {\bf B423} (1998) 64.

\bibitem{}
M. Garousi, "Super string Scattering from D-brane bound states", 
hep-th/9805078.


\bibitem{}
K. Kaviani, M.M. Sheikh-Jabbari, work in preparation.


\end{thebibliography}
\end{document}